\documentstyle[prd,aps,preprint,floats,amsfonts,amsbsy]{revtex}
\def\pacs#1{\par %
\bgroup
\hsize\columnwidth \parindent0pt
\if@twocolumn\else\leftskip=0.10753\textwidth \rightskip\leftskip\fi
\ifdim\prevdepth=-1000pt \prevdepth0pt\fi
\dimen0=-\prevdepth \advance\dimen0 by20pt\nointerlineskip
\vbox to28pt{\small\vrule height\dimen0 width0pt\relax\ifdraft#1\fi\vfill}%
\egroup
\if@twocolumn\vskip1pc\fi
\ifpreprintsty
\penalty10000\vfill
{\em GSI-Preprint 95-69}\\
Invited contribution to the Umezawa Memorial Volume of
Int.J.Mod.Phys. {\bf B}
\newpage
\fi
}
\begin{document}
\tighten
\def\vec#1{\bbox{#1}}
\def\hi{\vphantom{\int\limits_0^0}}
\def\lo{\vphantom{\int\limits}}
\title{Thermal Field Theory
  in Non-Equilibrium States}
\author{P.A.Henning}
\address{Theoretical Physics,
        Gesellschaft f\"ur Schwerionenforschung GSI\\
        P.O.Box 110552, D-64220 Darmstadt, Germany\\
        {\sc e-mail: P.Henning@gsi.de} }
\author{K.Nakamura}
\address{Division of Natural Science, Meiji University, Izumi Campus\\
        Eifuku, Suginami-ku, Tokyo 168 Japan\\
        {\sc e-mail: aa00001@isc.meiji.ac.jp} }
\author{Y.Yamanaka}
\address{Waseda University High School\\
        3-31-1 Kanishakuji, Nerima-ku, Tokyo 177 Japan\\
        {\sc e-mail: yamanaka@cfi.waseda.ac.jp}}
\maketitle
\begin{abstract}
Conventional transport theory is not really applicable to
non-equilibrium systems which exhibit strong quantum effects.
We present two different approaches to overcome this problem.
Firstly we point out how transport equations may be derived that
incorporate a nontrivial spectral function as a typical quantum effect,
and test this approach in a toy model of a strongly interacting
degenerate plasma. Secondly we explore a path to include
non-equilibrium effects into quantum field theory through
momentum mixing transformations in Fock space.
Although the two approaches are completely orthogonal, they
lead to the same coherent conclusion.\\[1cm]
Dedicated to the memory of Hiroomi Umezawa,\\
Friend and mentor,\\
Who had the vision to unify \\
Quantum field theory and statistical mechanics.\\
\end{abstract}
\section{Introduction}
In the past two decades the interest in non-equilibrium quantum
systems has grown tremendously. It is therefore natural, that
more and more examples have been found for which the applicability
of traditional {\em transport theory\/} is doubtful, i.e.,
macroscopic Boltzmann-equations or Vlasov-Uehling-Uhlenbeck equations fail
to describe the system adequately.
Indeed some of these examples, like dynamical features of the early universe,
collective effects in fusion plasmas, ultrashort phenomena in semi-conductors,
nuclear matter in ultrarelativistic heavy-ion collisions and
plasma drops composed of quarks and gluons have two things in
common that require a new paradigm for their theoretical understanding.

The first of these common factors is the (space and time)
density of interactions between the components of the system.
Some of them are interacting weakly in the sense of having a small
coupling constant, e.g., as between electrons and photons. However,
even for weak coupling, a system of {\em many\/} particles
experiences synergetic effects like {\em collective\/} participation
in movement. As a consequence, the systems listed above have
excitation spectra which differ significantly from those of rarefied gases
of free particles. The latter however is a basic requirement
for the applicability of traditional transport theory.

The second common factor is the importance of the
non-equilibrium aspect on the level of the system components:
Inhomogeneities occur on spatial and temporal scales that are
comparable to the intrinsic scales of the physical problem.
Particularly interesting examples are ultrarelativistic heavy-ion
collisions, e.g. between lead nuclei ${}^{208}$Pb at 160$\times$208
GeV laboratory energy. Here, time scales are fm/c and characteristic
sizes are a few fm -- not significantly different from the
diameter of the nucleons which compose the initial nuclei.
Such a significant difference however is another requirement
for the description of the system in terms of {\em macroscopic\/}
equations: Traditional transport theory is correct only to first
order of gradients in density, temperature etc.

We may summarize the two common aspects of the systems listed above
as {\bf 1.~}the breakdown of the quasi-particle picture, and
{\bf 2.~}the non-separation of space-time scales.
The necessary theoretical description therefore
must account for these two aspects.

With the present paper we wish to contribute to this new paradigm from
two sides: In section II we shall point out a way to incorporate
nontrivial excitation spectra into transport theory (which already
contains some non-equilibrium features), and in section III we will clean
a path for the inclusion of non-equilibrium effects (gradients
in external parameters) into quantum field theory.
These two completely different approaches
lead to the same consistent conclusion, summarized in
section IV.
\section{From standard to quantum transport theory}
We perform this step by studying the reverse direction, i.e.,
starting from a Schwinger-Dyson equation for the full
two-point function of a fermionic quantum field we derive
transport equations. In contrast to the standard treatment however, we
do {\em not\/} perform the usual quasi-particle approximation.

As has been pointed out by various authors, (see ref. \cite{LW87}
for an overview) the description of dynamical quantum phenomena
in a statistical ensemble necessitates a formalism with a doubled
Hilbert space.  For our purpose the
relevant content of this formalism is, that its two-point Green
functions are 2$\times$2 matrix-valued. We prefer the technically
simpler method of thermo field dynamics (TFD)\cite{Ubook},
but will keep our derivation sufficiently general to
perform similar calculations in the Schwinger-Keldysh,
or Closed-Time Path (CTP) formalism \cite{SKF}.

Within this matrix formulation, we consider
the Schwinger-Dyson equation for the full quark propagator
$
S =  S_0  +  S_0 \odot \Sigma \odot S
$, where
$S_0$ is the free and $S$ the full two-point Green function
of the quark field, $\Sigma$ is the full self energy
and the generalized product $\odot$ is
a matrix product (thermal and spinor indices) and an
integration (each of the matrices is a function of two space
coordinates). Throughout this paper we use the convention to write
space-time and momentum variables also as
lower indices, e.g. $\Sigma_{xy}\equiv \Sigma(x,y)$.

In the CTP formulation as well as in the $\alpha=1$ parameterization
of TFD \cite{h94rep}, the matrix elements of $S$, $S_0$ and $\Sigma$ obey
\begin{equation}\label{sme}
S^{11}_{(0)}+S^{22}_{(0)}=S^{12}_{(0)}+S^{21}_{(0)}
\;;\;\;\;\;\;\Sigma^{11}+\Sigma^{22}=-\Sigma^{12}-\Sigma^{21}
\;.\end{equation}
Therefore the four components of the Schwinger-Dyson equation are
not independent, the matrix equation can be simplified by a linear
transformation which one may
conveniently express as a matrix product \cite{RS86}.
It achieves a physical interpretation only in the TFD formalism,
see ref. \cite{h94rep}. The transformation matrices ${\cal B}$ are
\begin{equation}\label{lc}
{\cal B}(n) =
\left(\array{lr}(1 - n) &\; -n\\
                1     & 1\endarray\right)
\;,\end{equation}
depending on one parameter only. For example, the third term in the
Schwinger-Dyson equation becomes
\begin{equation}\label{qptp}
  {\cal B}(n)\,\tau_3\,S_0\odot\Sigma\odot S\,({\cal B}(n))^{-1}
  = \left({\array{lr}  S_0^R\odot\Sigma^R\odot S^R & \mbox{something} \\
                       & S_0^A\odot\Sigma^A\odot S^A \endarray}\right)
\;.\end{equation}
Here, $\tau_3 = \mbox{diag}(1,-1)$,
$\Sigma^{R,A}$ are the retarded and advanced full self energy function,
and $S^{R,A}$ are the retarded and advanced full propagator
(similarly for $S_0$)
\begin{eqnarray}\label{sra}\nonumber
\Sigma^R = \Sigma^{11}+\Sigma^{12}\;,\;\;\;\;
&\Sigma^A = \Sigma^{11}+\Sigma^{21}\\
S^R =  S^{11}-S^{12}\;,\;\;\;\; &S^A = S^{11}-S^{21}
\;.\end{eqnarray}
The diagonal elements of the transformed equation therefore
are {\em retarded\/} and {\em advanced\/} Schwinger-Dyson equation.
The off-diagonal element is a {\em transport equation\/}.

We now switch to the mixed
(or Wigner) representation of functions depending on two
space-time coordinates:
$
\tilde\Sigma_{XP} = \int\!\!d^4(x-y) \,
  \exp\left({\mathrm i} P_\mu (x-y)^\mu\right)\Sigma_{xy}
$ with $X = (x+y)/2$,
the $\tilde{\ }$-sign will be dropped henceforth.
The Wigner transform of the convolution $\Sigma\odot G$
is a nontrivial step: Formally it may be expressed as
a gradient expansion
\begin{equation}\label{gex}
 \int\!\!d^4(x-y) \;
  \exp\left({\mathrm i} P_\mu (x-y)^\mu\right)\; \Sigma_{xz}\odot G_{zy}
 = \exp\left(-{\mathrm i}\Diamond\right)\,\tilde\Sigma_{XP} \,
  \tilde{G}_{XP}
\;.\end{equation}
$\Diamond$ is a 2nd order differential operator acting on both
functions appearing behind it in the form of a Poisson bracket
$
\Diamond A_{XP} B_{XP}    =
  \frac{1}{2}\left(\partial_X A_{XP} \partial_P B_{XP}-
                            \partial_P A_{XP}\partial_X B_{XP}\right)
$
{}.
We will henceforth use the infinite-order differential operator
$
\exp(-{\mathrm i} \Diamond)=
\cos\Diamond-{\mathrm i}\sin\Diamond
$.
Propagator and self energy are split into real Dirac matrix valued
functions
\begin{equation}\label{split}
S^{R,A}_{XP} = G_{XP} \mp {\mathrm i} \pi {\cal A}_{XP}
\;\;\;\;\;
\Sigma^{R,A}_{XP} =
  \mbox{Re}\Sigma_{XP} \mp {\mathrm i}\pi \Gamma_{XP}
\;.\end{equation}
${\cal A}_{XP}$ is the generalized spectral function of
the quantum field.

Now consider the equations obtained by action of Dirac
differential operators (= {\em inverse free propagators\/}) on the
matrix-transformed Schwinger-Dyson equation\cite{h94gl3}. The
diagonal components are
\begin{eqnarray}\nonumber
&&\mbox{Tr}\left[\left(
   P^\mu\gamma_\mu- m  \right)  {\cal A}_{XP}\right] =
  \cos\Diamond\,\mbox{Tr}\left[
  \mbox{Re}\Sigma_{XP}\, {\cal A}_{XP}
                         + \Gamma_{XP} \, G_{XP}\right]\\
\label{k8c}
&&\mbox{Tr}\left[\left(
   P^\mu\gamma_\mu- m   \right)  G_{XP}\right] =
  \mbox{Tr}\left[1\right] +
  \cos\Diamond\, \mbox{Tr}\left[
  \mbox{Re}\Sigma_{XP} \, G_{XP}
                         -\pi^2\,\Gamma_{XP}\,{\cal A}_{XP}\right]
\;.\end{eqnarray}
Two important facts about these equations have to be emphasized.
First notice that these equations do not in general
admit a $\delta$-function solution
for ${\cal A}_{XP}$ even in zero order of $\Diamond$. In fact,
in contrast to other papers \cite{hm93} we find that
there is not such thing as a mass shell constraint
in quantum transport theory.

Secondly, the equations do not contain odd powers of
the differential operator $\Diamond$. This implies, that when truncating
the Schwinger-Dyson equation to first order in $\Diamond$
(the usual order for the approximations leading to
{\em kinetic\/} equations), the spectral function ${\cal A}_{XP}$
may still be obtained as the solution of an algebraic equation.
\subsection{Transport equation}
The off-diagonal component of the transformed Schwinger-Dyson equation
reads, after acting on it with the inverse free propagator
\cite{h94rep,h94gl3}
\begin{equation}\label{k5}
\widehat{S}^{-1}_0   S^K_{xy}     =  \Sigma^R_{xz} \odot S^K_{zy}
        -              \Sigma^K_{xz} \odot S^A_{zy}
\;,\end{equation}
with kinetic components
$S^K  = \left( 1- n\right)\,S^{12} + n\,S^{21}$
and
$\Sigma^K = \left( 1- n\right)\,\Sigma^{12} +
        n\,\Sigma^{21}$.
Inserting the real functions defined before, this leads to
a differential equation, which contains all the features
of our desired quantum transport theory \cite{h94rep,h94gl3}:
\begin{eqnarray}\label{tpe1} \nonumber
\mbox{Tr}\left[\left(
  \partial_X^\mu\gamma_\mu + 2 \sin\Diamond\;
  \mbox{Re}\Sigma_{XP}
   + \cos\Diamond\;2\pi\Gamma_{XP}
  \right)  S^K_{XP}\right]& = \\
   2{\mathrm i} \mbox{Tr}\left[
             {\mathrm i}\sin\Diamond\;\Sigma^K_{XP} \, G_{XP}
             - \cos\Diamond\;\Sigma^K_{XP} \,
    {\mathrm i}\pi{\cal A}_{XP}\right]&
\;.\end{eqnarray}
Note, that here even as well as odd powers of the operator
$\Diamond$ occur, and the solution in zero order $\Diamond$ is
not trivial. To see this more clearly, we {\em define\/} the generalized
covariant distribution function $N_{XP}$ through the equation
\begin{equation}\label{nde}
\left(1-N_{XP}\right)\,S_{XP}^{12} + N_{XP}\,S_{XP}^{21} =0
\,.\end{equation}
It may be proven easily from the Kubo-Martin-Schwinger boundary condition
\cite{KMS}, that with this definition of $N_{XP}$ one
reaches the proper equilibrium limit $N_{XP} \rightarrow n_F(E)$, where
$n_F(E)$ is the Fermi-Dirac equilibrium distribution function at
temperature $T$,
\begin{equation}
n_F(E) = (\mbox{e}^{ \beta (E-\mu)}+1)^{-1}
\;.\end{equation}
With this definition,
$
S_{XP}^K = 2\pi{\mathrm i}\,\left(N_{XP} - n\right)\,{\cal A}_{XP}
$, and therefore $N_{XP}$ is the
parameter which diagonalizes the the full non-equilibrium
matrix-valued propagator through the Bogoliubov matrix
${\cal B}$ from (\ref{lc})\cite{h94rep}:
\begin{equation}\label{b1def}
{\cal B}(N_{XP})\,\tau_3\,S_{XP}\,({\cal B}(N_{XP}))^{-1}=
\left({\array{rr} G_{XP}-{\mathrm i}\pi{\cal A}_{XP} & \\
 & G_{XP}+{\mathrm i}\pi{\cal A}_{XP} \endarray}\right)
\;.\end{equation}
Although one may use the generalized distribution function
directly in the above eq. (\ref{tpe1}), we wish to
compare this equation to {\em traditional transport
theory\/}. Hence, one more step has to be performed, which is the
consistent expansion to first order in the operator
$\Diamond$.

As outlined before, we may then use a spectral function which is the solution
of an algebraic equation. One may argue, that in non-equilibrium states
a spectral representation of the propagator does not exist
in general \cite{Ubook}, but one may still exploit
the fact that retarded and advanced propagator are
by definition analytical functions of the
energy parameter in the upper or lower complex energy half-plane:
\begin{equation}\label{rapf}
S^{R,A}(E,\vec{p},X)  = \mbox{Re}{G}_{XP} \mp \pi {\mathrm i}
  {\cal A}_{XP} =
  \int\limits_{-\infty}^{\infty}\!\!dE^\prime\;
  {\cal A}(E^\prime,\vec{p},X)\;
   \frac{1}{E-E^\prime\pm{\mathrm i}\epsilon}
\;,\end{equation}
which is nothing but the Wigner transform of
$
S^{R,A}_{xy} = \mp 2\pi{\mathrm i}\Theta\left(\pm(x_0-y_0)\right)
        {\cal A}_{xy}
$.

Inserting this into the quantum transport equation the yields,
correct to first order in $\Diamond$
(see refs. \cite{h94gl3,h94rep} for details):
\begin{eqnarray}
\nonumber
&&\mbox{Tr}\left[ {\cal A}_{XP} \left\{  \hi
\left( \lo P_\mu\gamma^\mu - m - \mbox{Re} \Sigma_{XP}
 \right), N_{XP} \right\}\right]\\
\nonumber
&&\;\;= {\mathrm i} \mbox{Tr}\left[\hi {\cal A}_{XP} \left( \lo
N_{XP} \Sigma^{21}_{XP} - \left( N_{XP}-1\right)
 \Sigma^{12}_{XP}\right) \right]\\
\nonumber
&&\;\; -{\mathrm i}
\int\limits_{-\infty}^0\!\!d\tau \int\!\frac{dE}{2\pi}\,
\sin(\tau E)\,\mbox{Tr}\left[ \left\{  \hi
{\cal A}(X;P_0+E,\vec{P}),\right.\right.\\
\label{tpe1a}
&&\left. \left.\left(\lo
N_{XP} \Sigma^{21}(t+\tau/2,\vec{X};P)
- \left(N_{XP}-1\right) \Sigma^{12}(t+\tau/2,\vec{X};P)\lo\right)\hi
\right\}_{N}\right]
\;.\end{eqnarray}
In this equation, $\left\{\lo\cdot,\cdot\right\}$ denotes the Poisson
bracket, the index $N$ means that the derivatives
are not acting on $N_{XP}$.

The integral over the history of our system
constitutes a {\em memory term\/} of this generalized transport equation.
The memory effect therefore can be attributed to the nonzero
spectral width of the physical excitations in our system.

However, even when we send this spectral width parameter to zero
in the above equation, a new effect remains. Using a simple
quasi-particle spectral function
\begin{eqnarray}\nonumber
{\cal A}_{XP} &=&
\left( \lo P_\mu\gamma^\mu - m - \mbox{Re} \Sigma_{XP}\right) \\
&&\delta\left(P_0-E(t,\vec{x},\vec{P})\right)
   \;\left|\frac{\partial ( (P^\mu-\mbox{Re} \Sigma^\mu_{XP})^2
    - (M-\mbox{Re} \Sigma^s_{XP})^2)}{
     \partial P_0}\right|^{-1}
\;,\end{eqnarray}
where $E(t,\vec{x},\vec{P})$ is the (space-time and momentum
dependent) generalized energy of the corresponding ``particle''-like
state, one obtains after integration over $P_0$
\begin{eqnarray}\nonumber
&&\left( \frac{\partial N(\vec{x},\vec{p},t)}{\partial t}
  +\frac{\partial E(t,\vec{x},\vec{p})}{\partial \vec{p}}
   \frac{\partial N(t,\vec{x},\vec{p})}{\partial \vec{x}}
  -\frac{\partial E(t,\vec{x},\vec{p})}{\partial \vec{x}}
   \frac{\partial N(t,\vec{x},\vec{p})}{\partial \vec{p}}\right) \\
\label{VL2}
&=&\mbox{St}\left[\lo N(t,\vec{x},\vec{p})\right] +
   \delta\mbox{St}\left[\lo N(t,\vec{x},\vec{p})\right]
\;.\end{eqnarray}
The l.h.s. of this equation is the Vlasov (or streaming) part of
the standard transport equation, it is free of dissipation.

The first piece on the r.h.s. is the standard ``collision integral''
of the kinetic equation, while $\delta\mbox{St}$ is
a {\em time-local\/} correction which now contains the
three-dimensional Poisson bracket
of the $\Sigma^{12}$, $\Sigma^{21}$ self energies with the
function $N(t,\vec{x},\vec{p})$ \cite{YN94}.
We will consider this equation in more detail in
section III of this paper.

Before {\em solving\/} the generalized transport equation
for a simple example, we would like to point out that with
our derivation we have indeed taken into account the
two topics mentioned in the introductory statements.
The breakdown of the quasi-particle approximation is accounted for
by the nontrivial spectral function, and the
non-separation of time-scales is accounted for by the
Poisson bracket on the r.h.s. of eq. (\ref{tpe1a}).
\subsection{Solution of the generalized transport equation}
\begin{figure}[t]
\vspace*{75mm}
\includegraphics{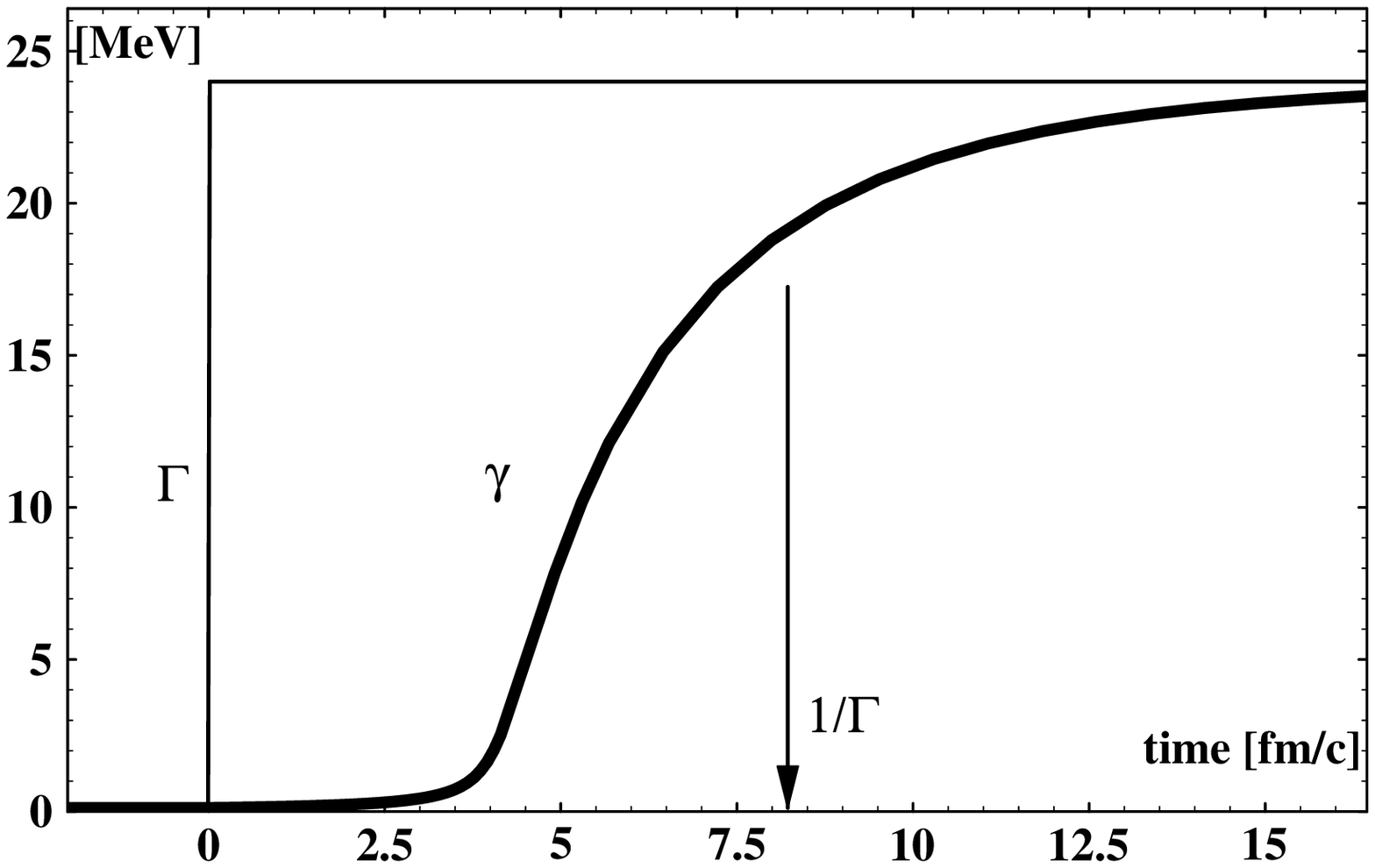}
\caption{Time dependent spectral width parameter
$\gamma_t$.\protect\newline
Parameters are $g$=0.12,  $T_i=$ 1 MeV, $T_f=$ 200 MeV,
$m=$ 10 MeV.\protect\newline
Thin line: $\Gamma_t$ from eq. (\ref{ss1}),
thick line: $\gamma_t$ from eq. (\ref{k9c}).
}
\vspace*{1mm}
\hrule
\end{figure}
\begin{figure}[t]
\vspace*{75mm}
\includegraphics{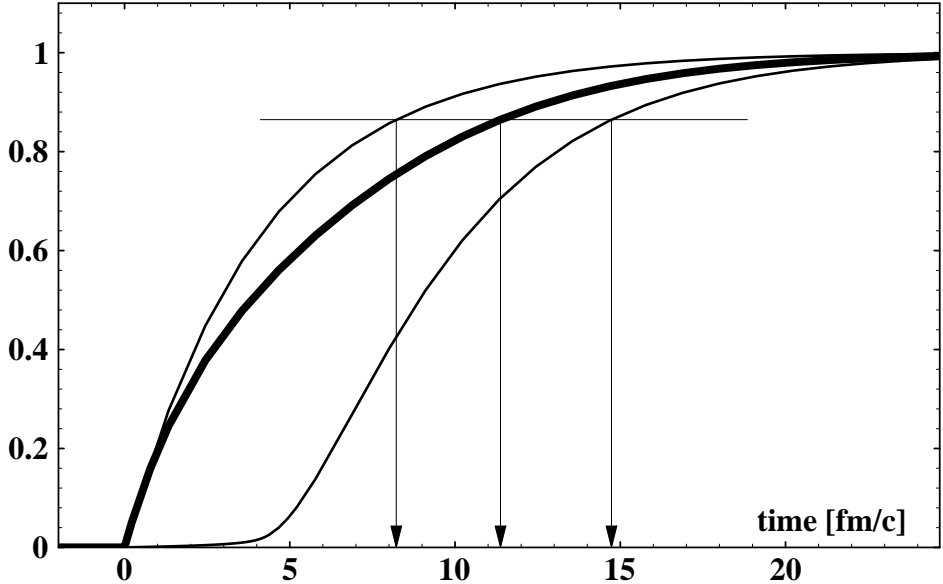}
\caption{Normalized time dependent fermionic distribution function for
 slow quarks.\protect\newline
Parameters as in Fig. 1;
thin lines: left $N^B_t/n_F(m,T_f)$ from the Boltzmann equation (\ref{tpe3}),
right $N_t/n_F(m,T_f)$ from the quantum transport equation (\ref{tpe2});
thick lines: $N^G_t/n_F(m,T_f)$
from the generalized kinetic equation (\ref{tpe5}).
}
\vspace*{1mm}
\hrule
\end{figure}
In the following, we will concentrate on the memory effects
hidden in eq. (\ref{tpe1a}). To isolate them from the
the gradients in the generalized collision integral
$\delta\mbox{St}\left[\lo N(t,\vec{x},\vec{p})\right]$
of eq. (\ref{VL2}), we consider a very simple model system,
tailored to mimic the thermalization of a plasma composed
of quarks and gluons. We cannot possibly explain all the
relevant physics in the present short paper, rather
refer to the literature on such systems \cite{QM95}.
In view of the previous derivations, we are interested
in a calculation of relaxation time scales for such
a plasma, and make some simple approximations:
\begin{enumerate}
\item We assume, that a gas of bosons (gluons) is
instantaneously heated to a very high temperature. In this
gas then eventually quark-antiquark pairs start to pop up,
until at the very end a thermal equilibrium in the sense of
a degenerate plasma is reached.
\item We assume, that the self energy function for the
quarks is dominated by gluonic contributions, and that
it does not depend on the energy of the quarks nor
on the space coordinates.
\item The gluon background is dominated by external conditions,
i.e., we neglect the back-reaction of quarks on the gluon distribution.
\item
We neglect the influence of anti-quarks in the spectral function.
This restriction is removed in an extended version of this application,
ref. \cite{h95neq}.
\end{enumerate}
We summarize these assumptions in the following ansatz for the
imaginary part of the self energy function and for the spectral
function of quarks:
\begin{equation}\label{ss1}
\Gamma_{XP}\equiv\Gamma_t = \gamma^0\;g T(t)\,= \gamma^0\;g\,
\left( T_i\,\Theta(-t)\,+\,T_f\,\Theta(t) \right)
\;,\end{equation}
\begin{equation} \label{fsf}
{\cal A}(t,E,\vec{p}) = \frac{\gamma^0}{\pi}
\frac{\gamma_t}{
  \left(E - \omega_t\right)^2 + \gamma_t^2}
\;.\end{equation}
Hence, we approximate the quark spectral function by two time-dependent
parameters $\omega_t$ and $\gamma_t$, which we may interpret as
effective energy and effective spectral width.
Arguments for the validity of this approach are given
in ref. \cite{h95neq}.

With the above spectral function the coupled system (\ref{k8c})
reduces to {\em a single\/} nonlinear equation
for $\gamma_t$, plus the condition
$
\omega^2_t = \omega^2_0 = \vec{p}^2 + m^2
$.
This latter condition is more complicated, when the anti-particle piece of
the spectral function is taken into account\cite{h95neq}.
The energy parameter is chosen as $E=\omega_0$, which yields instead of
eq. (\ref{k8c}) as the Schwinger-Dyson equation for the retarded
(or advanced) two-point function of the quarks:
\begin{equation}\label{k9c}
\gamma_t = g T_i + g (T_f - T_i)\,\Theta(t)\,
   \left(1-{\mathrm e}^{{\displaystyle -2 \gamma_t t}}\right)
\end{equation}
In Fig. 1, the solution of this equations is plotted in comparison
to the time dependent imaginary part of the self energy function
from eq. (\ref{ss1}). It is obvious, that the solution of the
nonlinear equation (\ref{k9c}) approaches
the imaginary part of the self energy function with a
characteristic delay time.
In ref.\cite{h95neq} it is discussed how this delay
time is calculated from the system parameters.

We now consider three different levels of transport theory
for this model, the corresponding generalized distribution functions
relabeled to $N_t$, $N^B_t$ and $N^G_t$.
First of all, due to the simplicity of
our ansatz we obtain as the full quantum transport equation (\ref{tpe1}):
\begin{equation}\label{tpe2}
\frac{d}{d t}N_t = -2 \,\gamma_t\left( N_t - n_F(m,T(t)) \right)
\;\end{equation}
with $T(t)$ as defined in eq. (\ref{ss1}).
This equation looks surprisingly similar to a kinetic equation in
relaxation time approach. However, this similarity is superficial:
The {\em kinetic\/} equation,
or Boltzmann equation, derived for our simple model system reads
\begin{equation}\label{tpe3}
\frac{d}{d t}N^B_t = -2 \,\Gamma_t\left( N^B_t - n_F(m,T(t)) \right)
\;.\end{equation}
Finally, the generalized transport equation (\ref{tpe1a})
is, correct up to first order in the gradients:
\begin{eqnarray}\nonumber
\frac{d}{d t} N^G_t & = &
 - 2 \Gamma_t\,\left(N^G_t - n_F(m,T(t))\right) \\
\nonumber
&& + 4\,t\,\Theta(t)\, \left(g\left(T_f-T_i\right)\right)^2\,
  \exp(-2 \Gamma_t t)\,\\
&&\;\;\;\;\;
  \left(N^G_t - \frac{n_F(m,T_f) T_f - n_F(m,T_i) T_i}{T_f - T_i}\right)
\label{tpe5}
\;.\end{eqnarray}
In Fig.2 we show the numerical solution for $N^G_t$, and compare
it to the Boltzmann solution $N^B_t$ as well as the full quantum
transport solution $N_t$.
This comparison of the three methods  shows,
that the full quantum transport equation
results in a {\em much \/} slower equilibration process
than the Boltzmann equation.
This result is in agreement with other attempts to solve
the quantum relaxation problem\cite{D84a,h93trans}: The quantum system
exhibits a memory, it behaves in an essentially non-Markovian way.

In particular, for the physical scenario studied here,
the time to reach 1-1/e${}^2\approx$ 86 \% of the equilibrium
quark occupation number is almost doubled
(14.7 fm/c as compared to 8.2 fm/c in the
Boltzmann case). We furthermore find, that with the generalized transport
equation one does at least partially describe the
memory effects in a quantum system (the characteristic time now is
11.4 fm/c).

Thus, although we have only used a toy model,
it might turn out that quantum effects (= memory as described in this
contribution) substantially hinder the thermalization of a
strongly interacting over long time scales.
A more thorough discussion of this physical result is carried out
in ref. \cite{h95neq}.
\section{From quantum field theory to inhomogeneous TFD}
In the following we describe a system
of charged bosons in a non-equilibrium state.
For simplicity we first consider a  space-inhomogeneous
situation, but neglect the spectral width of the bosons.
This view therefore is orthogonal to the
previous section, where we ended in considering a time-dependent
system with translational invariance.
While this gave us a window to isolate the memory effects
of quantum transport theory, we will isolate some
non-equilibrium quantum features in the present picture.

As we have pointed out, the description of non-equilibrium
systems requires to double the Hilbert space.
The reason is, that the occupation number, i.e., the interpretation of
a state as ``particle'' or ``hole''
(and hence its temporal boundary condition) may change from
point to point in order to ensure causality.

The proper formulation is achieved by expressing each
physical (causal) excitation operator as the superposition of an
operator evolving forward in time  and a backward evolving operator.
Such a superposition is known in the Liouville space, we refer
to the literature for an introduction \cite{S78,H86}.
Obviously, there is also a corresponding anti-causal (orthogonal) linear
combination. Denoting the causal
single-particle creation and annihilation operators
by $a_{ql}$, $a_{ql}^\dagger$ (the two different charges are distinguished
by a lower index $l=\pm$) and the
anti-causal operators with  $\widetilde{~}$-signs, one may conveniently
express this superposition as a matrix in three-momentum space
\cite{NUY92,NUY88}:
\begin{eqnarray}
\nonumber
\left({\array{r} a_{kl}(t)\\
          \widetilde{a}^\dagger_{kl}(t)\endarray}\right)\;=
& \displaystyle\int\!\!d^3\vec{q}\,
  \left({\cal B}^{-1}_l(t,\vec{q},\vec{k})\right)^\star
  \,\left({\array{r}\xi_{ql}\\
              \widetilde{\xi}^\#_{ql}\endarray}\right)
  \,{\mathrm e}^{-\mathrm{i} E_q t} \nonumber \\
\left({\array{r}a^\dagger_{kl}(t)\\
         -\widetilde{a}_{kl}(t)\endarray}\right)^T\;=
& \displaystyle\int\!\!d^3\vec{q}\,
  \left({\array{r}\xi^\#_{ql}\\
\label{mix1}
           -\widetilde{\xi}_{ql}\endarray}\right)^T\,
  {\cal B}_l(t,\vec{q},\vec{k})
  \,{\mathrm e}^{\mathrm{i} E_q t}
\;.\end{eqnarray}
$\vec{k}$ is the three-momentum of the modes, therefore in this notation
$a^\dagger_{k-}(t)$ creates a negatively charged
physical excitation with momentum $\vec{k}$,
while $a_{k+}(t)$ annihilates a positive charge.

The operator $\xi^\#_{ql}$ creates a mode with
momentum $\vec{q}$ and energy $E_q\equiv E(\vec{q})$, which
propagates forward in time. $\widetilde{\xi}_{ql}$ annihilates a state
with the same momentum and energy, but propagating backwards in time.
These two operators therefore may be combined in a Bogoliubov transformation.
with a $2\times 2$ matrix (compare to the fermion case, eq. (\ref{b1def})):
\begin{equation}\label{gb}
{\cal B}_l(t,\vec{q},\vec{k}) = \left( { \array{lr}
   \left(\delta^3(\vec{q}-\vec{k}) + N_l(t,\vec{q},\vec{k})\right)
            \;\;\;& -N_l(t,\vec{q},\vec{k}) \\
   -\delta^3(\vec{q}-\vec{k})     & \delta^3(\vec{q}-\vec{k})
   \endarray} \right)
\;.\end{equation}
For simplicity, let us choose a local equilibrium state, i.e.,
$N(t,\vec{q},\vec{k})$ is the Fourier transform of a space-local
Bose-Einstein distribution function
\begin{eqnarray}\nonumber
N_l(t,\vec{q},\vec{k})& = & \frac{1}{(2\pi)^3}\;
        \int\!\!d^3\vec{z}\,{\mathrm e}^{-\mathrm{i}
  (\vec{q}-\vec{k})\vec{z} }\,n_l(t,\vec{z},(\vec{q}+\vec{k})/2)\\
\label{nloc}
n_l(t,\vec{z},\vec{p}) & = &
  \left[{\mathrm e}^{\beta(t,\vec{z}) (E_p-\mu_l(t,\vec{z}))}-1\right]^{-1}
\;.\end{eqnarray}
With these creation and annihilation operators
we construct two mutually commuting complex scalar
fields $\phi_x$, $\widetilde{\phi}_x$
see ref. \cite{LW87,h94rep,h93trans} for details:
\begin{eqnarray}\label{bf1}
\phi_x & = &
   \int\!\! \frac{d^3\vec{k}}{\sqrt{(2\pi)^3}}
   \left( a^\dagger_{k-}(t)\,{\mathrm e}^{-\mathrm{i}\vec{k}\vec{x}} +
          a_{k+}(t)\,        {\mathrm e}^{ \mathrm{i}\vec{k}\vec{x}}\right)
                \nonumber \\
\widetilde{\phi}_x & = &
   \int\!\! \frac{d^3\vec{k}}{\sqrt{(2\pi)^3}}
   \left( \widetilde{a}^\dagger_{k-}(t)\,{\mathrm e}^{
  \mathrm{i}\vec{k}\vec{x}} +
          \widetilde{a}_{k+}(t)        \,{\mathrm e}^{
 -\mathrm{i}\vec{k}\vec{x}}
                \right)
\;.\end{eqnarray}
Each of these is a representation of the canonical
commutation relations, and they may be combined in a statistical doublet
$\Phi_x=\left(\phi_x,\widetilde{\phi}_x\right)^T$.

Before proceeding we would like to emphasize that
the above heuristic formulation has led us to the creation and annihilation
operators known from thermo field dynamics
(TFD, \cite{Ubook}). However, the construction of two commuting representations
in non-equilibrium field theory is also achieved in other formulations
\cite{LW87}, and their existence has been noticed independently
by several authors \cite{S78,H86}.
\subsection{Effective interaction and diffusion}
In this subsection we will rely more on results obtained in
the TFD formalism than in the other parts of this paper. However,
we believe that the approach described in the following is also valid
and useful for other methods used in thermal field theory.

First of all we note, that
due to the introduction of the momentum mixing terms
the equation of motion for the physical creation/annihilation operators
is
\begin{equation}\label{heis}
 {\mathrm i} \frac{\partial}{\partial t}\,
 \left({\array{r} a_{kl}(t)\\
         \widetilde{a}^\dagger_{kl}(t)\endarray}\right)\;=\;
 \left[\left({\array{r} a_{kl}(t)\\
         \widetilde{a}^\dagger_{kl}(t)\endarray}\right),
                \widehat{H}+\widehat{{\cal Q}}\right]
\;,\end{equation}
with a ``bare Hamiltonian''
\begin{equation}\label{hnn}
  \widehat{H} = \sum\limits_{l=\pm}
\int\!\!d^3\vec{k}\,E_k\,
        \left({\array{r} a^\dagger_{kl}(t)\\
         -\widetilde{a}_{kl}(t)\endarray}\right)^T\,
        \left({\array{r} a_{kl}(t)\\
         \widetilde{a}^\dagger_{kl}(t)\endarray}\right)
=\sum\limits_{l=\pm} \int\!\!d^3\vec{k}\,E_k\,
  \left( a^\dagger_{kl}(t) a_{kl}(t) -
  \widetilde{a}^\dagger_{kl}(t) \widetilde{a}_{kl}(t) \right)
\;\end{equation}
and a second part that vanishes for homogeneous systems:
\begin{equation}\label{hen}
\widehat{{\cal Q}} = \sum\limits_{l=\pm}
\int\!\!d^3\vec{k}\,d^3\vec{q}\,
        \left({\array{r} a^\dagger_{kl}(t)\\
         -\widetilde{a}_{kl}(t)\endarray}\right)^T\,
        \left({\array{lr} 1 & 1 \\ 1 & 1 \endarray}\right)
        \,  \left( -{\mathrm i}\partial_t+E_k-E_q\right)
        \,N_l(t,\vec{k},\vec{q})\,
        \left({\array{r} a_{ql}(t)\\
         \widetilde{a}^\dagger_{ql}(t)\endarray}\right)\,
\;.\end{equation}
Consequently, by introducing the above linear combinations for the
{\em physical\/} creation and annihilation operators, the
physical fields are no longer free. The time evolution acquires
a term mixing causal and anti-causal field.

Let us emphasize at this point, that one should
think of $\widehat{H} + \widehat{Q}$ as the {\em Liouville\/} operator
of our quantum system, which is the generator of the time evolution
for the density matrix. Its spectral properties
then are immediately obvious \cite{S78,H86,NRT83,L88}, e.g., it
is not bounded from below.

The mixing part $\widehat{{\cal Q}}$ of the full Liouvillean vanishes when
$N_l(t,\vec{k},\vec{q})$ is time independent and
proportional to $\delta(\vec{k}-\vec{q})$, which
according to (\ref{nloc}) is the case when $n_l(t,\vec{z},\vec{p})$
does not depend on the space-time coordinates $(t,\vec{z})$.
Consequently $\widehat{{\cal Q}}$
couples the systems to {\em gradients\/} in the
function $n_l(t,\vec{z},\vec{p})$.

To see this more explicitly, we henceforth introduce
a ``trivial'' dispersion relation $E_q=\sqrt{\vec{q}^2+m^2}$
and obtain
\begin{equation}
\left(E_k-E_q\right)\,N_l(t,\vec{k},\vec{q})\,
= -{\mathrm i}\,\frac{\vec{Q}}{2E_Q(2\pi)^3}\,\int\!\!d^3\vec{z}
\,{\mathrm e}^{-\mathrm{i}
  (\vec{q}-\vec{k})\vec{z} }\,
  \nabla_z\, n_l(t,\vec{z},\vec{Q})
 +{\cal O}(\nabla_z^2)
\;,\end{equation}
where $\vec{Q}=(\vec{q}+\vec{k})/2$.
One may argue, that this is an ad-hoc introduction of a coupling into
the system dynamics. However, this formulation has a deeper foundation:
Consider for a moment an equilibrium system, because then
the Liouville operator is identical to eq. (\ref{hnn}) and invariant under
the thermal Bogoliubov transformation as
$
\left( a^\dagger_k a_k - \widetilde{a}^\dagger_k \widetilde{a}_k
\right) =
\left( \xi^\#_k \xi_k - \widetilde{\xi}^\#_k \widetilde{\xi}_k
\right)
$.
This invariance constitutes of a continuous symplectic symmetry of the
Liouville space -- and such a global symmetry can be made {\em local\/}:
We search for those operators $\xi$, which
diagonalize the Liouville operator {\em locally\/} in space and time.
Such a requirement connects the parameter $n_l(t,\vec{z},\vec{p})$
of the Bogoliubov transformation with the system dynamics.

Therefore, this process of gauging the symplectic symmetry
is entirely equivalent to the
space-time local Bogoliubov transformation {\em diagonalizing\/}
the single-particle propagator that was carried out in sect. II.

On the other hand, it is well known that gauging of a symmetry automatically
introduces the coupling of the field $\phi_x$ to gradients in an
external scalar field. Here, this external scalar field is the
distribution function ``field'' $n_l(t,\vec{z},\vec{p})$,
it is classical and does not possess any dynamical feature.
To summarize these arguments: The new gradient terms in the time evolution
occur, because we are implementing {\em causal\/} boundary conditions
locally in space and time.

Let us now turn to the generalized transport equation used in the previous
section. Rewritten for bosons, with an effective
energy parameter $E$ that is independent of $\vec{x}$
it reads
\begin{equation}\label{VL2b}
\left(  \frac{\partial n_l(t,\vec{x},\vec{k})}{\partial t}
  + \frac{\partial E(\vec{k})}{\partial \vec{k}}
   \frac{\partial n_l(t,\vec{x},\vec{k})}{\partial \vec{x}}\right) \\
=\mbox{St}\left[\lo n_l(t,\vec{x},\vec{k})\right] +
   \delta\mbox{St}\left[\lo n_l(t,\vec{x},\vec{k})\right]
\;.\end{equation}
According to our derivation in section II we know that
$\delta\mbox{St}$
contains a three-dimensional Poisson bracket of $n$ and some
interaction probability. Obviously this implies that
\begin{equation}
 \int\!\!\frac{d^3\vec{k}}{(2\pi)^3}\int\!\!d^3\vec{x}\;
 \delta\mbox{St}\left[\lo n_l(t,\vec{x},\vec{k})\right]
 =0
\;.\end{equation}
and therefore we may write the momentum integral as the
three-dimensional divergence of a current:
\begin{equation}
 \int\!\!\frac{d^3\vec{k}}{(2\pi)^3}\;
 \delta\mbox{St}\left[\lo n_l(t,\vec{x},\vec{k})\right]
 = -\nabla_x\,\vec{J}_l^\delta(t,\vec{x})
\;\end{equation}
with $\nabla_x \equiv \partial/\partial\vec{x}$.
We use the fact, that the momentum integral over the
``standard'' collision term $\mbox{St}\left[\lo n(t,\vec{x},\vec{k})\right]$
is zero \cite{L90}, and therefore obtain for the integrated transport equation
\begin{equation}\label{cc}
\partial_t n_l(t,\vec{x}) + \nabla_x\,
\left(\vec{j}_l(t,\vec{x})+\vec{J}_l^\delta(t,\vec{x})\right) = 0
\;.\end{equation}
This is the expression for {\em current conservation\/}, with
\begin{equation}
\vec{j}_l(t,\vec{x})= \int\!\!\frac{d^3\vec{k}}{(2\pi)^3}\;
  \frac{\vec{k}}{E_k}\,n_l(t,\vec{x},\vec{k})
\;\end{equation}
as the convective part of this current.
Eq. (\ref{cc}) implies, that in our boson gas a current
$\vec{J}_l^\delta(t,\vec{x})$ arises even
for zero ``convection'' $\vec{j}_l(t,\vec{x})$: An inhomogeneous
temperature distribution gives rise to diffusion.
For small gradients of the temperature distribution
the diffusion current $\vec{J}_l^\delta(t,\vec{x})$ will be
proportional to  $\nabla_x\,n_l(t,\vec{x})$. The
momentum integrated transport equation then is nothing but Fick's law.

More detailed derivations within a non-relativistic field model,
as well as exemplaric calculations of particle flow and energy flow
may be found in ref. \cite{YN95}.
\subsection{Calculation of the Diffusion Coefficient}
Having outlined the diffusion problem in the previous
subsection, we need to generalize the momentum mixing to
excitations with continuous mass spectrum.
This amounts to an approximation of the fully interacting quantum fields
by {\em generalized free fields\/} \cite{L88} as
already discussed in section II, and will certainly
complicate matters very much.

On the other hand we have
outlined above, that only the spatial gradient terms enter
into the diffusion problem. Hence, for the following
we may neglect the time dependence of the mixing parameter
$n$. Instead of eq. (\ref{mix1}) we therefore write
\begin{eqnarray}
\nonumber
\left({\array{r} a_{kl}(t)\\
          \widetilde{a}^\dagger_{kl}(t)\endarray}\right)\;=
& \int\limits_0^\infty\!\!dE\,\int\!\!d^3\vec{q}
  \;{\cal A}^{1/2}_l(E,\vec{k})\,
  \left({\cal B}^{-1}_l(E,\vec{q},\vec{k})\right)^\star
  \,\left({\array{r}\xi_{Eql}\\
              \widetilde{\xi}^\#_{Eql}\endarray}\right)
  \,{\mathrm e}^{-\mathrm{i} Et} \nonumber \\
\left({\array{r}a^\dagger_{kl}(t)\\
         -\widetilde{a}_{kl}(t)\endarray}\right)^T\;=
& \int\limits_0^\infty\!\!dE\,\int\!\!d^3\vec{q}
  \;{\cal A}^{1/2}_l(E,\vec{k})\,
  \left({\array{r}\xi^\#_{Eql}\\
           -\widetilde{\xi}_{Eql}\endarray}\right)^T\,
  {\cal B}_l(E,\vec{q},\vec{k})
  \,{\mathrm e}^{\mathrm{i} E t}
\;,\end{eqnarray}
where the $2\times 2$ Bogoliubov matrices ${\cal B}$ have the same form as
already given in eq. (\ref{gb}), but $N_l$ is replaced by
\begin{eqnarray}\nonumber
N_l(E,\vec{q},\vec{k})& = & \frac{1}{(2\pi)^3}\;
        \int\!\!d^3\vec{z}\,{\mathrm e}^{-\mathrm{i}
  (\vec{q}-\vec{k})\vec{z} }\,n_l(E,\vec{z})\\
\label{nloc2}
n_l(E,\vec{z}) & = &  \left[\lo
  {\mathrm e}^{\beta(\vec{z}) (E-\mu_l(\vec{z}))}-1\right]^{-1}
\;.\end{eqnarray}
The $\xi$-operators have commutation relations
\begin{equation}\label{difc}
\left[\xi_{Ekl},\xi^\#_{E^\prime k^\prime l^\prime}\right]=
  \delta_{ll^\prime}\,
  \delta(E-E^\prime)\,
\delta^3(\vec{k}-\vec{k}^\prime)
\;.\end{equation}
Similar relations hold for the $\widetilde{\xi}$ operators,
all other commutators vanish, see \cite{L88}.

The weight functions ${\cal A}_l(E,\vec{k})$ are positive and have
support only for positive energies, their normalization is
\begin{equation}\label{norm}
\int\limits_0^\infty\!\!dE\,E\, {\cal A}_l(E,\vec{k}) =\frac{1}{2}\,,
\;\;\;\;\;\;
\int\limits_0^\infty\!\!dE \, {\cal A}_l(E,\vec{k}) =Z_{kl}
\;.\end{equation}
The principles of this expansion have been discussed in ref. \cite{L88},
its generalization to non-equilibrium states was introduced
in ref. \cite{h94rep}. For equilibrium states the combination
$
{\cal A}_B(E,\vec{k}) = {\cal A}_+(E,\vec{k})\Theta(E) -
                        {\cal A}_-(-E,-\vec{k})\Theta(-E)
$
is the spectral function of the boson field $\phi_x$
and the limit of free particles with mass  $m$ is recovered when
$
{\cal A}_B(E,\vec{k}) \longrightarrow
\mbox{sign}(E)\,  \delta(E^2 -\vec{k}^2 -m^2)
=\mbox{sign}(E)\,
  \delta(E^2 -\omega_k^2)
$.
As we have seen in the previous section, the concept of a local
spectral function may be applied to non-equilibrium states up to first
order in the gradient operator $\Diamond$.

Using these relations we are then able to express the current
{\em response\/} of the quantum field to the gradients
in the distribution function by using the Heisenberg equation of
motion for time-dependent operators, eq. (\ref{heis}), to obtain
\begin{equation}\label{td}
\vec{J}^\delta(t,\vec{x}) = {\mathrm i} \int\limits_{t_0}^t\!\!d\tau
  \left\langle \hi
  \left[\widehat{\vec{j}}(t-\tau,\vec{x}),\widehat{\cal Q}\right]\right\rangle
\;\end{equation}
where $\widehat{\vec{j}}(x) = {\mathrm i}(
\phi_x^\dagger\nabla_x\phi_x-\phi_x\nabla_x\phi^\dagger_x)
$ is the current operator.

As has been pointed out in \cite{h94rep,h93trans},
the $i$-th vector component of the $l$-charged currents
generated by the inhomogeneity of the system is
\begin{eqnarray}\nonumber
\vec{J}^{\delta(i)}_l(t,\vec{x})&=\;\displaystyle
   2\pi\,\int\!\!\frac{d^3\vec{Q}}{(2\pi)^3}\,
   \frac{\vec{Q}^{(i)}}{2 Z^2_{Ql}}&
   \int\!\!dE\,\left({\cal A}_l(E,\vec{Q})\right)^2
   \times \\
 &&\int\!\!dE^\prime\,E^\prime\,\left\{
   \frac{\partial n_B^l(E^\prime,\vec{Q},\vec{x})}{\partial \vec{x}^{(j)}}
   \frac{\partial {\cal A}_l(E^\prime,\vec{Q})}{\partial \vec{Q}^{(j)}}
   \right\}
\label{curl}
\;.\end{eqnarray}
In principle, this equation may be used to calculate
the diffusion constant for interacting boson fields.
The result of such a calculation made for pions coupled to
nuclear matter has led to results which are  bigger
than the semi-classical transport coefficients \cite{HST84,L90}.
by factors of 10--100 \cite{h93trans,h94rep}.
Since the transport coefficients
are proportional to the relaxation time of the system, this
corresponds to the result obtained in section II:
Quantum effects lead to a slowdown of the relaxation process.

In order to establish our approach more safely, we now insert a
boson spectral function with a constant spectral width. However,
in contrast to the fermionic case discussed in section II, we may then
no longer neglect the anti-particles. Hence we use
\begin{equation}
{\cal A}_B(E,\vec{k}) = \frac{1}{\pi}
\frac{2 E \gamma_B}{
  \left(E^2 - E^2_k - \gamma^2_B\right)^2 + 4 E^2 \gamma_B^2}
\;.\end{equation}
Similar to the fermionic spectral function
(\ref{fsf}), one obtains this form by summing over four simple
poles in the complex energy plane, each with the same distance from the
real $E$-axis and with the same residue. A short calculation then
gives the approximate results
\begin{eqnarray}\nonumber
\frac{1}{Z^2_{Ql}}\,\int\!\!dE\,\left({\cal A}_l(E,\vec{Q})\right)^2
 &=&\frac{1}{2\pi\gamma_B} + \frac{2}{\pi^2 E_Q} +{\cal O}(\gamma_B)\\
\int\!\!dE^\prime\,E^\prime\,
   \frac{\partial n_l(E^\prime,\vec{Q},\vec{x})}{\partial \vec{x}^{(j)}}
   \frac{\partial {\cal A}_l(E^\prime,\vec{Q})}{\partial \vec{Q}^{(j)}}
&=&\frac{\vec{Q}^{(j)}}{|\vec{Q}|}\,\frac{\partial}{\partial |\vec{Q}|}\,
\frac{\partial}{\partial \vec{x}^{(j)}}\,n_l(E_Q,\vec{x})
\;,\end{eqnarray}
such that an integration by parts gives, up to first order in
the gradients of the distribution function,
\begin{equation}
\vec{J}^{\delta}_l(\vec{x}) = -\left(
 \frac{1}{2\gamma_B} + {\cal O}(1)\right)\,\nabla_x\,n_l(t,\vec{x})
 = - D\,\nabla_x\,n_l(t,\vec{x})
\;.\end{equation}
The diffusion coefficient $D$
therefore diverges when $\gamma_B\rightarrow 0$. In
traditional calculations of transport coefficients
this diffusion coefficient is obtained
as $D=\tau \left\langle \vec{v}^2\right\rangle/3$, i.e.,
as the product of relaxation time and square average of the particle
velocity \cite{L90}. As we have shown before, the relaxation
time of a non-equilibrium quantum system is infinite if the particles
do not have a spectral width -- and therefore our result has
the correct free-particle limit.

For more detailed considerations however, we would have to define a
model for an interacting system, which e.g. yields
the spectral width parameter as function of the system characteristics
(temperature, chemical potential etc.). This would overstress
the goal set with the present paper.
\section{Conclusion}
With the present paper we have demonstrated, how one may
apply modern techniques of ``thermal'' field theory to
non-equilibrium states. This has become important because
of the quantum character of non-equilibrium systems studied
today: A definite breakdown of the quasi-particle approximation
and the emergence of space-time inhomogeneities on scales
comparable to those of system components are the reasons
for the non-applicability of ``standard'' transport theory.

We have started with the derivation of a generalized transport equation,
which accounts for the breakdown of the quasi-particle
approximation. In Fig. 2 we have compared the solution of this equation
with the full Schwinger-Dyson equation as well as with
standard transport theory. We find, that the nontrivial spectral
function one encounters when going beyond the quasi-particle picture
leads to strong {\em memory effects\/} in transport phenomena.
In particular, we studied a system which exhibits a characteristic delay time.
The memory effects are partially taken into account by the
generalized transport equation -- but it turns out, that
gradients in the external parameters are equally important to
describe the physical systems listed in the introduction to this
work.

In the second major part of this work we therefore pointed out,
how one may incorporate gradient effects into ``thermal'' field theory.
This is achieved by introducing a mixing of operators in
momentum space. Firstly we have shown that such an approach
indeed leads to relaxation currents in inhomogeneous systems.
Secondly we have used the mixing transformation to achieve
a linear response result for transport coefficients.

This transport coefficient calculation was not pursued numerically.
However, we may rely on previous numerical estimates of
transport coefficients achieved with this method \cite{h94rep}.
Those numerical estimates give a result very similar to the one
obtained in the quantum transport model of section II:
Quantum systems are relaxing {\em slower\/} than anticipated from
standard transport theory.

Hence, the two approaches we have demonstrated in the present paper
lead to a coherent result, which is also in agreement with
more traditional calculations from statistical
mechanics. We therefore believe that a clear path exists towards
an improved treatment of quantum systems in non-equilibrium states,
and we express our gratitude to our teacher Hiroomi Umezawa
for leading us on this path.


\begin{thebibliography}{99}
\bibitem{LW87}{
    N.P.Landsman and Ch.G.van Weert,
    Phys.Rep. {\bf 145} (1987) 141}
\bibitem{Ubook}{
    L.Leplae, H.Umezawa and F.Mancini,
    Phys.Rep. {\bf 10} (1974) 151;\\
    T.Arimitsu and H.Umezawa,
    Prog.Theor.Phys. {\bf 77} (1987) 32 and 53;\\
    H.Umezawa,\\
    { Advanced Field Theory: Micro, Macro and Thermal Physics}\\
    (American Institute of Physics, 1993)}
\bibitem{SKF}{
    J.Schwinger,
    J.Math.Phys. {\bf 2} (1961) 407;\\
    L.V.Keldysh, Zh.Exsp.Teor.Fiz. {\bf 47} (1964) 1515 and
    JETP {\bf 20} (1965) 1018}
\bibitem{h94rep}{
    P.A.Henning, Phys.Rep. {\bf 253} (1995) 235}
\bibitem{RS86}{
    J.Rammer and H.Smith,
    Rev.Mod.Phys. {\bf 58} (1986) 323}
\bibitem{h94gl3}{
    P.A.Henning,
    Nucl.Phys. {\bf A582} (1995) 633
    (Erratum: {\bf A586} (1995) 777)}
\bibitem{hm93}{
    S.Mrowczynski and U.Heinz,  Ann.Phys. {\bf 229} (1994) 1}
\bibitem{KMS}{
    R. Kubo,
    J.Phys.Soc. Japan {\bf 12} (1957) 570;\\
    C.Martin and J.Schwinger,
    Phys.Rev. {\bf 115} (1959) 1342}
\bibitem{YN94}{
    Y.Yamanaka and K.Nakamura,
    Mod.Phys.Lett {\bf A 9} (1994) 2879}
\bibitem{QM95}{
   {\em Proc. of Quark Matter '95}, Monterey, 1995 (to be
   published);\\
   Quark Matter 93 Proceedings, Nucl. Phys.
   {\bf A 566,} 1c (1994) }
\bibitem{h95neq}{
   P.A.Henning, E.Quack and P.Zhuang,\\
   {\em Thermalization of a Quark-Gluon Plasma}\\
   GSI-Preprint 1995 in preparation, also published in the Proceedings of the\\
     4-th International Workshop on Thermal
     Field Theories and Their Applications, Dalian}
\bibitem{D84a}{
    P.Danielewicz,
    Ann.Phys. {\bf 152} (1984)  239 and 305}
\bibitem{h93trans}{
    P.A.Henning, Nucl.Phys. {\bf A567} (1994) 844}
\bibitem{S78}{
    M.Schmutz,
    Z.Physik {\bf B30} (1978) 97}
\bibitem{H86}{
    L. van Hove,
    Phys.Rep. {\bf 137} (1986) 11}
\bibitem{NUY92}{
    K.Nakamura, H.Umezawa and Y.Yamanaka,
    Mod.Phys.Lett {\bf A 7} (1992) 3583}
\bibitem{NUY88}{
    K.Nakamura, H.Umezawa and Y.Yamanaka,
    Physica {\bf A 150} (1988) 118}
\bibitem{NRT83}{
    H.Narnhofer, M.Requardt and W.Thirring,\\
    Commun.Math.Phys. {\bf 92} (1983) 247}
\bibitem{L88}{
    N.P.Landsman,
    Ann.Phys. {\bf 186 } (1988) 141}
\bibitem{L90}{
  R.L. Liboff, { Kinetic Theory}
  (Prentice-Hall, Englewood Cliffs, NJ 1990)}
\bibitem{YN95}{ 
    Y.Yamanaka and K.Nakamura,\\
    {\em Diffusion in inhomogeneous TFD}
     to be published in the Proceedings of the\\
     4-th International Workshop on Thermal
     Field Theories and Their Applications, Dalian}
\bibitem{HST84}{
    A.Hosoya, M.Sakagami and M.Takao,
    Ann.Phys. {\bf 154} (1984) 229}
\end{thebibliography}
\end{document}